\documentclass[11pt,a4paper]{article}
\usepackage{cite,palatino,epsfig,amsmath,amssymb}
\usepackage{graphics}
\usepackage{epsf}
\usepackage{rotating}
\usepackage{psfrag}
\usepackage[varg]{txfonts}
\usepackage{dcolumn}
\usepackage{feynmp} 
\usepackage{lcreport} 

\usepackage[]{datetime}

\usepackage{lineno}

\input axodraw.sty
\input rotate
\newcommand{\comment}[1]{}

\newcommand{\beq}{\begin{equation}}
\newcommand{\eeq}{\end{equation}}
\newcommand{\bea}{\begin{eqnarray}}
\newcommand{\eea}{\end{eqnarray}}
\newcommand{\barr}{\begin{array}}
\newcommand{\earr}{\end{array}}

\newcommand{\gsim}
{{\;\raise0.3ex\hbox{$>$\kern-0.75em\raise-1.1ex\hbox{$\sim$}}\;}}

\newcommand{\lsim}
{{\;\raise0.3ex\hbox{$<$\kern-0.75em\raise-1.1ex\hbox{$\sim$}}\;}}

\def\gev{\, {\rm GeV}}
\def\tev{\, {\rm TeV}}

\allowdisplaybreaks

\long\def\symbolfootnote[#1]#2{\begingroup%
\def\thefootnote{\fnsymbol{footnote}}\footnote[#1]{#2}\endgroup}

\def\cut#1{}

\begin{document}









\setcounter{page}{1}


\thispagestyle{empty}

\begin{flushright}
LC-REP-2012-071,~ILC ESD-2012/4,~CLIC-Note-949 (July 30, 2012)
\end{flushright}

\begin{center}

{\Large\sc \bf\boldmath The Physics Case for an $\epem$ Linear Collider}

\vspace*{.7cm}

James E. Brau${}^a$,
Rohini M. Godbole${}^b$, 
Francois R. Le Diberder${}^{c}$,
M.A. Thomson${}^d$,\\
Harry Weerts${}^e$,
Georg Weiglein${}^f$,
James D. Wells${}^g$,
Hitoshi Yamamoto${}^h$
\bigskip

{\it A Report Commissioned by the Linear Collider Community\symbolfootnote[2]{See Addendum for this committee's origin and charge. The committee also wishes to express thanks to the many reviewers of earlier drafts of this report whose input has been very helpful.}}

\begin{flushleft}
${}^{(a)}$Center for High Energy Physics, University of Oregon, USA;
${}^{(b)}$Centre for High Energy Physics, Indian Institute of Science, Bangalore, India; 
${}^{(c)}$Laboratoire de l'Acc\'el\'erateur Lin\'eaire, IN2P3/CNRS et Universit\'e Paris-Sud, France; 
${}^{(d)}$Cavendish Laboratory, University of Cambridge, UK; 
${}^{(e)}$Argonne National Laboratory, Argonne, USA; 
${}^{(f)}$DESY, Hamburg, Germany; 
${}^{(g)}$CERN, Geneva, Switzerland;
${}^{(h)}$Tohoku University, Japan
\end{flushleft}

\end{center}



\section{Introduction}
\label{section:intro}

The physics motivation for an $\epem$ linear collider (LC) has been
studied in detail for more than 20 years~\cite{Abe:2001gc}-\cite{Abe:2001wn}. These studies 
have provided a compelling case for a LC as the next collider at the energy frontier.  
The unique strengths of a LC stem from the clean experimental
environment arising from $\epem$ collisions. In particular, the
centre-of-mass energy and initial-state polarisations~\cite{MoortgatPick:2005cw} are precisely known 
and can be adjusted, and backgrounds are many orders of magnitude lower than the QCD backgrounds that challenge hadron collider environments.  
The low backgrounds  permit trigger-free readout,
and the measurements and searches for new phenomena are unbiased and comprehensive.
Full event reconstruction is possible.
These favourable experimental conditions will enable the LC to measure the properties of 
physics at the TeV scale with unprecedented precision and complementarity to the LHC. 

Thanks to the extraordinary achievements of the LHC machine and of the ATLAS and CMS experiments, our field witnessed a deep revolution in the middle of 2012: the discovery of a new boson. The observation at the LHC of this new particle compatible with a light Higgs boson strengthens the physics case for a LC even more.

The main goals of the LC physics programme are: 
\begin{itemize}
\item precise
measurements of the properties of the Higgs sector; 
\item precise measurements of the interactions of top quarks, gauge bosons, and new particles;
\item searches for physics beyond the Standard Model (SM), 
where, in particular,
the discovery reach of the LC can significantly exceed that
of the LHC  for the pair-production of colour-neutral states; and 
\item sensitivity to new physics 
through tree-level or quantum effects in high-precision observables.
\end{itemize}
The complementarity of the LC and LHC has been established over many years
by a dedicated worldwide collaborative effort~\cite{Weiglein:2004hn}. It has been shown in many contexts 
that for new particles found at the LHC, the LC will be essential in determining the properties of these new particles and unraveling the underlying structure of the new physics.

The development of the SM was a triumph for modern science. 
The experimental confirmation of the
${\rm SU(3)}_{\rm C}\times {\rm SU(2)}_{\rm L}\times{\rm U(1)}_{\rm Y}$ 
gauge structure of the SM and the precise measurement of its parameters were achieved through  a
combination of analyses of data from $\epem$ and hadron colliders and from
deep-inelastic lepton-nucleon scattering.
These precision measurements are compatible with 
the minimal Brout-Englert-Higgs mechanism of Electroweak Symmetry Breaking (EWSB), 
through which the masses of all the known fundamental particles are generated. 
Within the SM the measurements of electroweak precision
observables show a pronounced preference for a light Higgs boson, below about 150~GeV.

The observation of a new particle compatible with a Higgs boson of mass $\sim 125$\,GeV is a major breakthrough in particle physics. It represents one of the most significant discoveries of modern science.
Given the far-reaching consequences for our
understanding of the fundamental structure of matter and the basic laws
of nature, it is of the highest priority to probe the properties 
of this particle with a comprehensive set of high-precision measurements to address such questions as: 
\begin{itemize}
\item What are the couplings of this particle to other known elementary particles? Is its coupling to each particle proportional to that particle's mass, as required in the SM by the Higgs mechanism?
\item What are the mass, width, spin and CP properties of this particle?
\item What is the value of the particle's self-coupling? Is this consistent with the expectation  from the symmetry-breaking potential? 
\item Is this particle a single, fundamental scalar as in the SM, or is it part of a larger structure? Is it part of a model with additional scalar doublets? Or, could it be a composite state, bound by new interactions?
\item Does this particle couple to new particles with no other couplings to the SM? Is the particle mixed with new scalars of exotic origin, for example, the radion of extra-dimensional models?
\end{itemize}
The LC provides a unique opportunity to study Higgs properties
with sufficient precision to answer these fundamental questions. The large numbers of 
Higgs bosons that would be produced at a LC, between $10^5$ and $10^6$
depending on centre-of-mass energy and integrated luminosity, and the clean 
final states mean that a LC can be considered as a Higgs factory where the properties of the Higgs boson
can be studied in great detail. In particular, a LC provides the
possibility of model-independent measurements of the Higgs couplings to the gauge bosons and 
fermions at the few percent level. 

Whilst the discovery of a signal compatible with a 
Higgs boson at the LHC represents a breakthrough in particle physics, 
it should be kept in mind that the minimal EWSB theory of the SM without other dynamical mechanisms has theoretical shortcomings, and a richer and 
more complex structure is generally favoured. Most of the ideas for physics beyond the SM (BSM) 
are driven by the need to achieve a deeper understanding 
of the EWSB mechanism. Furthermore, the presence of non-baryonic dark matter
in the cosmos is an experimentally established fact that implies BSM physics. 
To date, no clear sign of BSM physics has emerged from LHC data. 
For new states that are colour-neutral,
a LC provides excellent sensitivity for direct discovery via pair
production. This complements the search reach of the LHC, where the
highest sensitivity is achieved for BSM coloured states. 
Should the two machines be operating concurrently, the LC results could even provide feed-back to the
LHC experiments and vice versa.

The flexibility of the LC will give rise to a rich physics programme which could consist of
i) a low-energy phase with $\sqrt{s}$ in the range of $250-500\, {\rm GeV}$, enabling the study of 
$\Zzero\Higgs$, $\Qt \AQt$, $\Higgs\Higgs\Zzero$ and $\Qt\AQt\Higgs$ thresholds,
and ii) a high-energy phase with $\sqrt{s} > 500$\,GeV allowing a high statistics study of the Higgs boson through the 
$\Wboson\Wboson$ fusion process and allowing access to rarer Higgs production processes such as
$\epem\rightarrow\Higgs\Higgs\nue\Anue$. The choice of the centre-of-mass energy range for the higher energy operation
would be determined by the BSM physics scale, where the flexibility in energy of a LC would allow the threshold behaviour for any
new physics process to be mapped out in detail.  While this document focuses on the minimal LC programme, there are a number of 
optional phases of LC operation, like  GigaZ, which is a high-luminosity $\Zzero$-factory, and $\electron\electron$, $\electron\photon$ and $\photon\photon$ configurations.  

Two options for a future $\epem$ LC have been developed, with different main linac acceleration schemes. 
The International Linear Collider (ILC) uses superconducting RF, 
whereas the Compact Linear Collider (CLIC) uses a separate drive beam to provide the accelerating power.
The ILC technology is mature and provides an option for a Higgs and top factory to be constructed on a relatively 
short timescale. 
The CLIC technology provides the potential to reach higher centre-of-mass energies, but it requires further development. 
In recent years there has been extensive collaboration between ILC and CLIC physicists with the goal of realising a 
LC as the next major new facility. Furthermore, the ILC and CLIC are being organised under the same formal 
worldwide body, the Linear Collider Board (LCB), reporting directly to ICFA. 
The strong accelerator development programme is complemented by an active theory and 
experimental community working on the 
physics and detectors for a future LC. These studies have resulted in detailed designs for the detectors at a LC, and, based
on detailed simulation studies, have provided a clear demonstration that the LC physics goals can be achieved. 



A comprehensive review of LC physics has been given in the Physics volume of the ILC RDR 
report~\cite{Djouadi:2007ik}, with extensions to higher energies in the CLIC CDR~\cite{Linssen:2012hp}.   More recently, many important measurements at the ILC and CLIC have been simulated with fully realised model detectors~\cite{Linssen:2012hp,Aihara:2009ad,Abe:2010aa,Barklow:2012}.  Finally, new reports on LC physics have attempted to bring the discussion of the LC capabilities~\cite{Linssen:2012hp,Barklow:2012} up to date in relation to recent results from the LHC. The main results from these physics studies are summarised below 
within the context of the results that have been obtained at the LHC up
to now and with a view also to the possible progress from the running of
the LHC during the next years. Unless otherwise stated, the discussion
refers generically to a Linear Collider (LC) rather than to the specific
realisations ILC or CLIC.



\section{Higgs Physics and Electroweak Symmetry Breaking}

In the SM, the Higgs boson plays a special role. The Higgs mechanism
is responsible for 
electroweak symmetry breaking and accounts for the generation of the masses of all 
the other elementary particles. 
In order to distinguish a SM
Higgs from possible alternative scenarios, it is necessary to measure precisely its  
couplings to the gauge bosons, to the fermions, and to itself.
Furthermore, the spin and the CP-properties of the new state need
to be determined, and it must be clarified whether there is more than one
physical Higgs boson. 
At the LHC ratios of the Higgs couplings to different particles can be measured for a subset of the possible
decays. Earlier LHC studies~\cite{Gianotti:2005} suggest that 
even with 3000\,fb$^{-1}$ of data the precision achievable on these ratios remains somewhat limited,
$\Gamma_\Wboson/\Gamma_\Zzero \sim 10\%$,   
$\Gamma_\Wboson/\Gamma_\Qb \sim 25\%$ and $\Gamma_\Wboson/\Gamma_\tauon \sim 30\%$.
It should be anticipated that with real data in hand LHC experiments likely will perform  better than initially projected.
At a LC, the precisions achievable are of the order of a few percent, 
and a wider range of decay channels can be studied. 

A LC is the only place where model-independent 
measurements of the Higgs boson couplings can be made, including to invisible final states. 
A number of these measurements are unique to a LC, and the precision achievable for extracting parameters 
significantly surpasses 
that anticipated at the LHC. The LC measurements would establish whether the Higgs boson has the properties predicted by the 
SM, or is part of an extended Higgs sector such as in SUSY models, 
or whether it has a
completely different physical origin which would be the case for a
composite Higgs.

\subsection{Higgs Production at a Linear Collider}

\begin{figure}[t]
\begin{center}
\raisebox{0.0cm}{\includegraphics[width=0.25\textwidth]{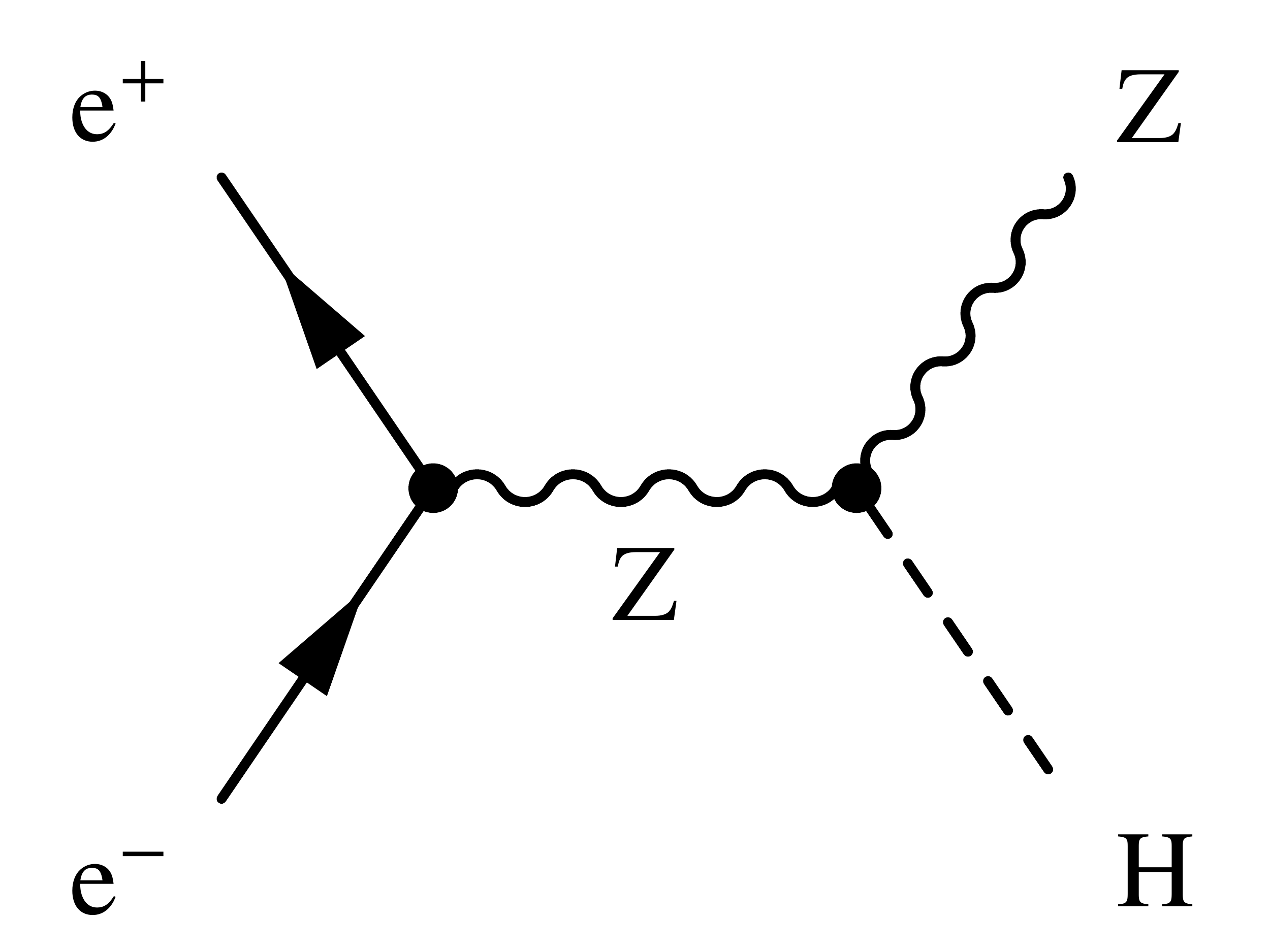}}
\raisebox{0.0cm}{\includegraphics[width=0.25\textwidth]{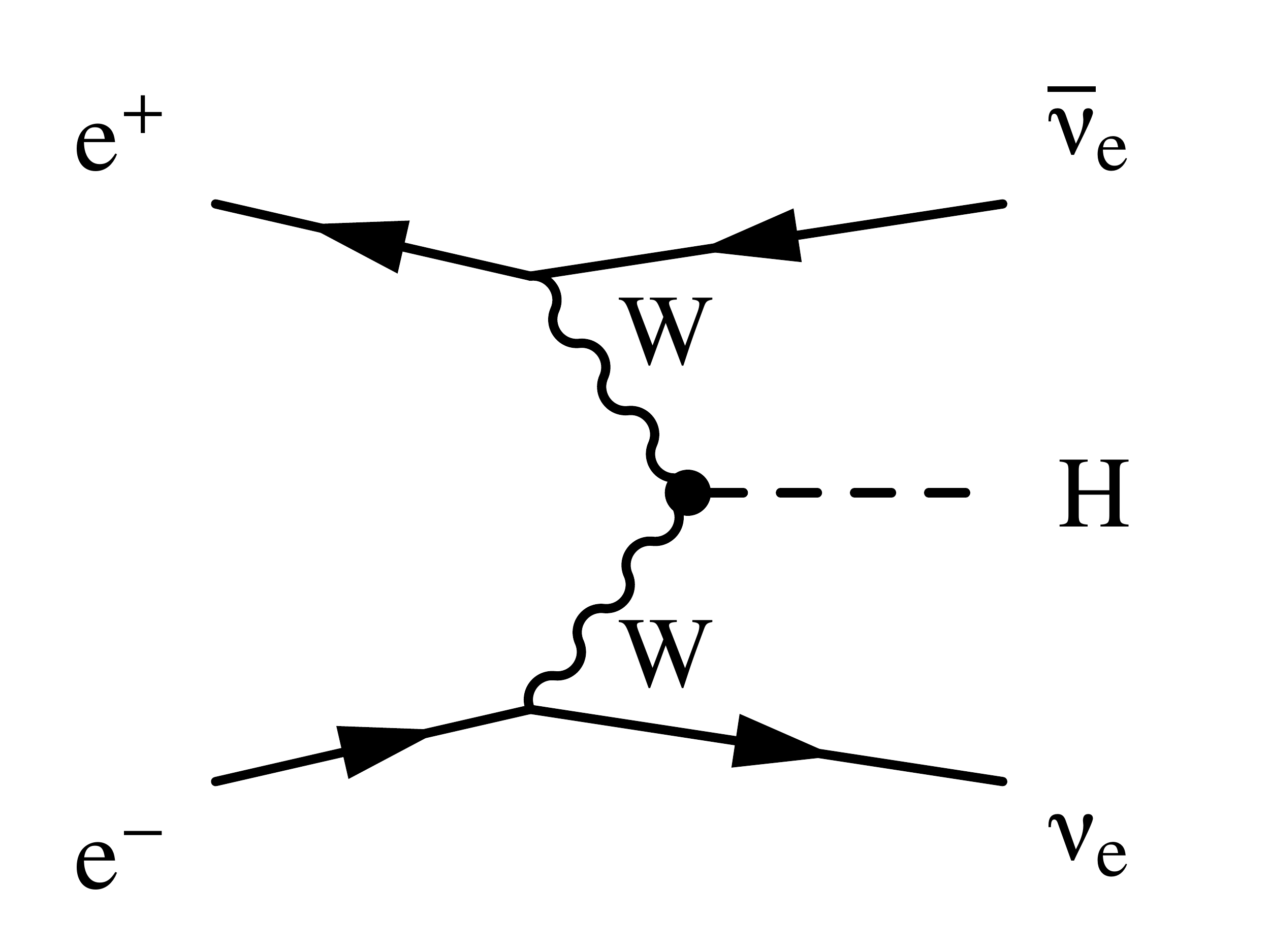}}
\caption{The two main Higgs production processes at a LC.Ê
Ê\label{fig:higgs:eezh}}
Ê \end{center}
\end{figure}

At a LC, the main Higgs production channels are through the Higgs-strahlung 
and vector boson fusion processes (see Figure~\ref{fig:higgs:eezh}).  
At relatively low centre-of-mass energies the Higgs-strahlung process,
$\epem\rightarrow\Zzero\Higgs$, dominates,
with a peak cross section at approximately 30\,GeV above the $\Zzero\Higgs$ production threshold. At higher centre-of-mass energies,  the $\Wboson\Wboson$ fusion process $\epem\rightarrow\Higgs\nue\Anue$ becomes increasingly important. For a 
Higgs boson mass of $\sim 125\gev$
the fusion process dominates above $\roots\sim500$\,GeV. The $\Wboson\Wboson$ fusion cross section
increases 
with $\roots$, allowing large samples of Higgs bosons to be studied at a TeV-scale LC. 
The $\Zzero\Zzero$ fusion process $\epem\rightarrow\Higgs\epem$ has a cross section that is approximately an order of magnitude smaller than the $\Wboson\Wboson$ fusion process. Table~\ref{tab:higgs:events} compares the expected number of $\Zzero\Higgs$ and $\Higgs\nue\Anue$ events at the
main centre-of-mass energies considered in the ILC and CLIC studies. Even at the lowest LC energies considered,
large samples of Higgs bosons can be accumulated.  In addition to the main Higgs production processes, rarer processes such as
$\epem\rightarrow\Qt\AQt\Higgs$, $\epem\rightarrow\Zzero\Higgs\Higgs$ and $\epem\rightarrow\Higgs\Higgs\nue\Anue$ provide access to the top 
quark Yukawa coupling and the Higgs trilinear self-coupling.

\begin{table}[t]
\begin{center}
\begin{tabular}{l|cccccc}
                                                                & 250\,GeV             & 350\,GeV      & 500\,GeV       & 1\,TeV            & 1.5\,TeV & 3\,TeV\\ \hline
$\sigma(\epem\rightarrow\Zzero\Higgs)$      &  240\,fb                & 129\,fb          &  57\,fb           &  13\,fb             & 6\,fb       & 1\,fb     \\        
$\sigma(\epem\rightarrow\Higgs\nue\Anue)$ &  8\,fb                   & 30\,fb            &  75\,fb           & 210\,fb            & 309\,fb    & 484\,fb     \\         
 Int. $\cal{L}$                                            &  250\,fb$^{-1}$       & 350\,fb$^{-1}$ & 500\,fb$^{-1}$& 1000\,fb$^{-1}$ & 1500\,fb$^{-1}$ & 2000\,fb$^{-1}$\\
 \# $\Zzero\Higgs$ events                         &  60,000                 &   45,500         &  28,500          &   13,000          & 7,500 & 2,000 \\ 
  \# $\Higgs\nue\Anue$ events                   &   2,000                 &   10,500         &  37,500          &   210,000         & 460,000  &  970,000  \\                          
\end{tabular}
\end{center}
\caption{The leading-order Higgs unpolarised cross sections for the Higgs-strahlung and $\Wboson\Wboson$-fusion processes at various centre-of-mass energies 
for $m_\Higgs=125$\,GeV. Also listed is the expected number of events accounting for the anticipated luminosities obtainable within 5 years of initial operation at each energy.  
\label{tab:higgs:events}}
\end{table}

\subsection{\boldmath Higgs Coupling Measurements at $\roots<500$ GeV}

\begin{figure}[t]
\begin{center}
 \includegraphics[width=0.50\linewidth]{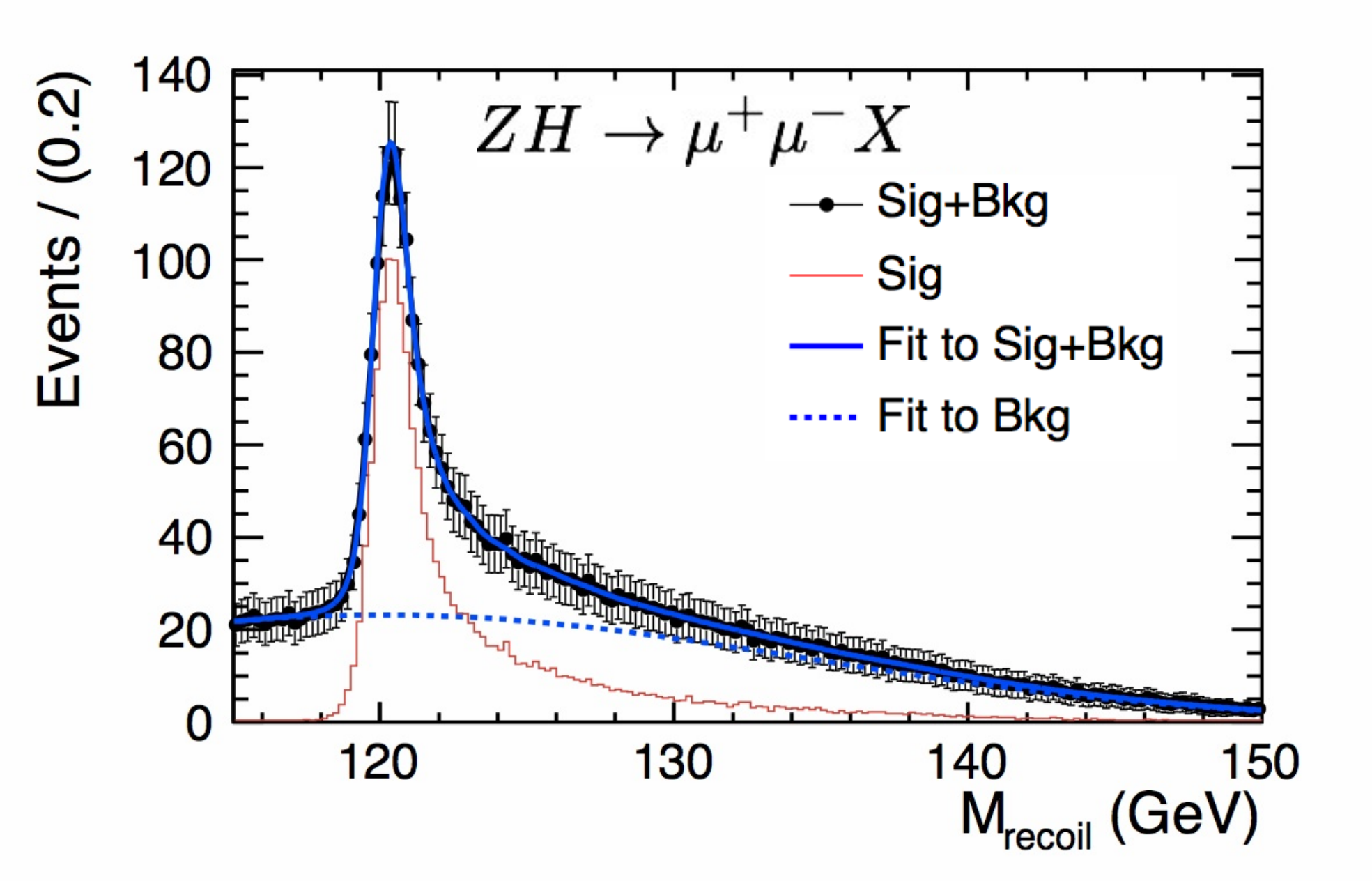}
 \caption{The recoil mass distribution for  $\epem\rightarrow\Zzero\Higgs\rightarrow\mpmm\Higgs$ events with $m_\Higgs=120\,\GeV$ in the ILD detector concept at the ILC\cite{Abe:2010aa}.
               The numbers of events correspond to 250\,fb$^{-1}$ at
$\roots=250\,\GeV$, and the error bars show the expected statistical uncertainties on the individual points.
 \label{fig:higgs:recoil}}
\end{center}
\end{figure}

The Higgs-strahlung process provides the opportunity to study the couplings of the Higgs boson in a 
{\it model-independent} manner. This is unique to a LC. The 
clean experimental environment, and the relatively low SM cross sections for 
background processes, 
allow  $\epem\rightarrow\Zzero\Higgs$ events to be selected based on the identification of two opposite 
charged leptons with invariant mass consistent with $\mZ$. The remainder of the event, i.e. the Higgs decay, is
not considered in the event selection. For example, Figure~\ref{fig:higgs:recoil} shows the simulated invariant mass distribution of the system recoiling against identified 
$\Zzero\rightarrow\mpmm$ decays at a LC for $\roots = 250$\,GeV.  A clear peak at the generated Higgs mass of $\mH=120$\,GeV is observed. Because only the properties of the di-lepton system are used in the selection, this method provides an absolute measurement of the Higgs-strahlung cross section, regardless of the Higgs boson decay modes; it would be equally valid if the Higgs boson decayed to invisible final states.  
Hence a model-independent measurement of the coupling $g_{\Higgs\Zzero\Zzero}$ can be made. 
With a dedicated analysis using also the hadronic decays of the Z the
sensitivity to invisible decay modes can be improved very significantly
as compared to the fully model-independent analysis. The LC provides in
fact a unique sensitivity to
invisible decay modes of the Higgs boson, extending down to a branching
ratio into invisible states as low as $1\%$.
The precisions achievable 
on the Higgs-strahlung cross section and the coupling $g_{\Higgs\Zzero\Zzero}$ are shown in Table~\ref{tab:higgs:zh}
for  $m_\Higgs=120$\,GeV. 

\begin{table}[t]
\begin{center}
\begin{tabular}{l|cc}
$\roots$                                      & 250\,GeV & 350\,GeV \\ \hline  
 Int. $\cal{L}$                                  &  250\,fb$^{-1}$  & 350\,fb$^{-1}$ \\
 $\Delta(\sigma)/\sigma$                               & 3\,\%       &  4\,\% \\
 $\Delta(g_{\Higgs\Zzero\Zzero})/g_{\Higgs\Zzero\Zzero}$ & {1.5\,\%}  & { 2\,\%} \\                             
\end{tabular}
\end{center}
\caption{Precision measurements of the Higgs coupling to the $\Zzero$ at $\roots=250$\,\GeV and
                 $\roots=350\,\GeV$ based  on full 
               simulation studies with $m_\Higgs=120\,\GeV$. Results from \cite{Abe:2010aa} and follow-up studies.
\label{tab:higgs:zh}}
\end{table}

The recoil mass study provides an absolute measurement of the total $\Zzero\Higgs$ production
cross section and therefore the total number of Higgs bosons produced would be known with a statistical precision of $3-4\,\%$.
The systematic uncertainties from the knowledge of the integrated luminosity and event selection are expected to be significantly smaller.
Subsequently, by identifying the individual final states for different Higgs and $\Zzero$ decay modes, absolute measurements of the Higgs boson 
branching fractions can be made. 
High flavour-tagging efficiencies are achievable and the $\Higgs\rightarrow\Qb\AQb$ and $\Higgs\rightarrow\Qc\AQc$ decays can be separated. Neglecting the Higgs decays into light quarks, one can also infer the branching ratio of
  $\Higgs\rightarrow \mathrm{g}\mathrm{g}$.
 Table~\ref{tab:higgs:brs} summarises the branching
 fraction precisions achievable at a LC  operating at either 250\,GeV or 350\,GeV where model-independent measurements of the Higgs boson couplings 
 to the $\Qb$-quark, $\Qc$-quark, $\tauon$-lepton, $\Wboson$-boson and $\Zzero$-boson can be made to better than 5\,\%.  

Preliminary results of ongoing studies confirm that a precision of $\Delta g_{\Qt\Qt\Higgs}/g_{\Qt\Qt\Higgs}\sim 10\%$ can be achieved, even near threshold at $500\gev$ with 1~ab${}^{-1}$, thanks to the factor of two enhancement of the QCD-induced bound-state effect. The measurement, which is made difficult by a very large $t\bar t$ background, relies on the foreseen performances of the LC detectors. Furthermore, $\Delta g_{H\gamma\gamma}/g_{H\gamma\gamma}$ can be measured at $\sim 5\%$ precision at a 500~GeV LC with 500~fb${}^{-1}$ of integrated luminosity.

\begin{table}[t]
\begin{center}
\begin{tabular}{l|ccccc|cccc|c|}
                                                                                & 250/350\,GeV &500\,GeV${}^\dagger$ & 3\,TeV &\phantom{----}&                 & 250/350\,GeV  & 500\,GeV${}^\dagger$ & 3\,TeV  \\ \cline{1-4} \cline{6-9} 
$\sigma\times Br(\Higgs\rightarrow\Qb\Qb)$   & $1.0/1.0\,\%$ & $0.6\,\%$   & $0.2\,\%$  &                           & $g_{\Higgs\Qb\Qb}$ & $1.6/1.4\,\%$ & $?$ & 2\,\% \\   
$\sigma\times Br(\Higgs\rightarrow\Qc\Qc)$    & $7/6\,\%$ &    $4\,\%$  & $3\,\%$     &                          & $g_{\Higgs\Qc\Qc}$  &    $4/3\,\%$ &  $2\,\%$  & 2\,\% \\
$\sigma\times Br(\Higgs\rightarrow\tauon\tauon)$  &  $6^*/6\,\%$  & $5\,\%$       & $?$ &                         & $g_{\Higgs\tauon\tauon}$  & $3^*/3\,\%$ & $2.5\,\%$  & $?$ \\
$\sigma\times Br(\Higgs\rightarrow\Wboson\Wboson)$ &  $8/6\,\%$  &   $3\,\%$    & $?$  &             & $g_{\Higgs\Wboson\Wboson}$ &  $4/3\,\%$ & $1.4\,\%$  & $<2\,\%$ \\     
$\sigma\times Br(\Higgs\rightarrow\muon\muon)$   &  $-/-$   &    $?$     & $15\,\%$ &                       & $g_{\Higgs\muon\muon}$  & $-/-$  &  $-$  & $7.5\,\%$ \\   
$\sigma\times Br(\Higgs\rightarrow \mathrm{g}\mathrm{g})$ & $9/7\,\%$ &   $5\,\%$  & $?$ & & $\frac{g_{\Higgs\Wboson\Wboson}}{g_{\Higgs\Zzero\Zzero}}$  & $?/?$ &  $?$  &$<1\,\%^*$     \\
  &  & & &  & $g_{\Higgs\Qt\Qt}$ & $-/-$ & $15\,\%$ & $?$ \\
 \end{tabular}
\end{center}
\caption{The precision on the Higgs branching ratios and couplings obtainable from studies of the Higgs-strahlung process at a LC operating at
               either $\roots=250$\,GeV,  $\roots=350$\,GeV and $\roots=500$\,GeV.
               The dagger on the 500\, GeV columns indicates that the quoted numbers are based on projections to be updated in~\cite{Barklow:2012}.
               The uncertainties on the
               couplings include the uncertainties on $g_{\Higgs\Zzero\Zzero}$ obtained from the absolute measurement of the $\Zzero\Higgs$ cross section. 
               Also shown are the precisions achievable from the $\Wboson\Wboson$ fusion process at a LC operating at 3\,TeV.
               The numbers marked with asterisk are estimates, all other numbers come from full 
               simulation studies with $m_\Higgs=120\,\GeV$. The question marks indicate that the results of ongoing studies are not yet available. 
               In all cases the luminosities assumed are those given in Table~\ref{tab:higgs:events}.
\label{tab:higgs:brs}}
\end{table}

\subsection{\boldmath Higgs Coupling Measurements at $\roots \ge 500$ GeV}

The large samples of events from both WW and $\Zzero\Zzero$ fusion processes would lead to a measurement
of the relative couplings of the Higgs boson to the $\Wboson$  and $\Zzero$ at the 1\,\% level.
This would provide a strong test of the SM prediction $g_{\Higgs\Wboson\Wboson} / g_{\Higgs\Zzero\Zzero}  =  \cos^2 \theta_W$.

The ability for clean flavour tagging combined with the large samples of $\Wboson\Wboson$ fusion events
allows the production rate of $\epem\rightarrow\Higgs\nue\Anue\rightarrow\Qb\AQb\nue\Anue$ to be determined with a precision
of better than 1\,\%.  Furthermore, the couplings to the fermions can be measured 
more precisely at high energies, even when accounting for the uncertainties on the production process. For example,
Table~\ref{tab:higgs:brs} shows the precision on the branching ratio obtained from full simulation studies 
as presented in~\cite{Linssen:2012hp}. The uncertainties of the  
Higgs couplings can be obtained by combining the high-energy results with those from the Higgs-strahlung process. The high statistics Higgs samples would 
allow for very precise measurements of relative branching ratios. For example, a LC operating at 3\,TeV would give a 
statistical precision of 1.5\,\% on $g_{\Higgs\Qc\Qc}/g_{\Higgs\Qb\Qb}$. 

\subsection{Higgs Self-Coupling}              

In the SM, the Higgs boson originates from a doublet of complex scalar fields described by the potential
\begin{equation*} 
       V(\phi) = \mu^2\phi^\dagger\phi + \lambda(\phi^\dagger\phi)^2 \,.
\end{equation*}
After spontaneous symmetry breaking, this form of the potential gives rise to a triple Higgs coupling of strength proportional to $\lambda v$, where $v$ is the 
vacuum expectation value of the Higgs potential. The measurement of the strength of the Higgs trilinear self-coupling therefore provides direct access to
the quartic potential coupling $\lambda$ assumed in the Higgs mechanism. This measurement is therefore
crucial for experimentally establishing the Higgs mechanism. 
For a low-mass Higgs boson, the measurement of the Higgs boson self-coupling at the LHC will be extremely challenging 
even with 3000\,fb$^{-1}$ of data.  
At a LC, the Higgs self-coupling can be measured through the $\epem\rightarrow\Zzero\Higgs\Higgs$
and $\epem\rightarrow\Higgs\Higgs\nue\Anue$
processes~\cite{AguilarSaavedra:2001rg}. 
The precision achievable is currently being studied for the $\epem\rightarrow\Zzero\Higgs\Higgs$ 
process at $\roots=500\,\GeV$ and for the $\epem\rightarrow\Higgs\Higgs\nue\Anue$ process at $\roots> 1$\,TeV.
Given the complexity of the final state and the smallness of the cross sections, these studies are being performed with a full simulation of the LC
detector concepts. The preliminary results 
indicate that a precision of about $20\,\%$ on $\lambda$ could be achieved, with the greatest sensitivity coming from   $\epem\rightarrow\Higgs\Higgs\nue\Anue$.

\subsection{Total Higgs Width}

For Higgs boson masses below 125 GeV, the total Higgs decay width in the SM ($\Gamma_\Higgs$) is less than 5 MeV and 
cannot be measured directly. Nevertheless, at a LC it can be determined from the relationship
between the total and partial decay widths, for example
\begin{equation*}
    \Gamma_H = {\Gamma(\Higgs\rightarrow \Wboson\Wboson^*)}/{Br(\Higgs \rightarrow \Wboson\Wboson^*)}\,.
\end{equation*}
Here
$\Gamma(\Higgs \rightarrow \Wboson\Wboson^*)$ can be determined from
the measurement of the $\Higgs\Wboson\Wboson$ coupling obtained from the fusion process $\epem\rightarrow \Higgs \nue\Anue$.  
When combined with the direct measurement of $Br(\Higgs \rightarrow \Wboson\Wboson^*)$, this allows the
Higgs width to be inferred. 
A precision on the total decay width of the Higgs boson 
of about 5\% at $\sqrt{s} = 500$\,GeV can be reached. This improves to better than 4\,\% at 1\,TeV.

\subsection{Impact of the Precision Measurements of the Higgs Couplings}

\begin{figure}[t]
\begin{center}
 \includegraphics[width=0.75\linewidth]{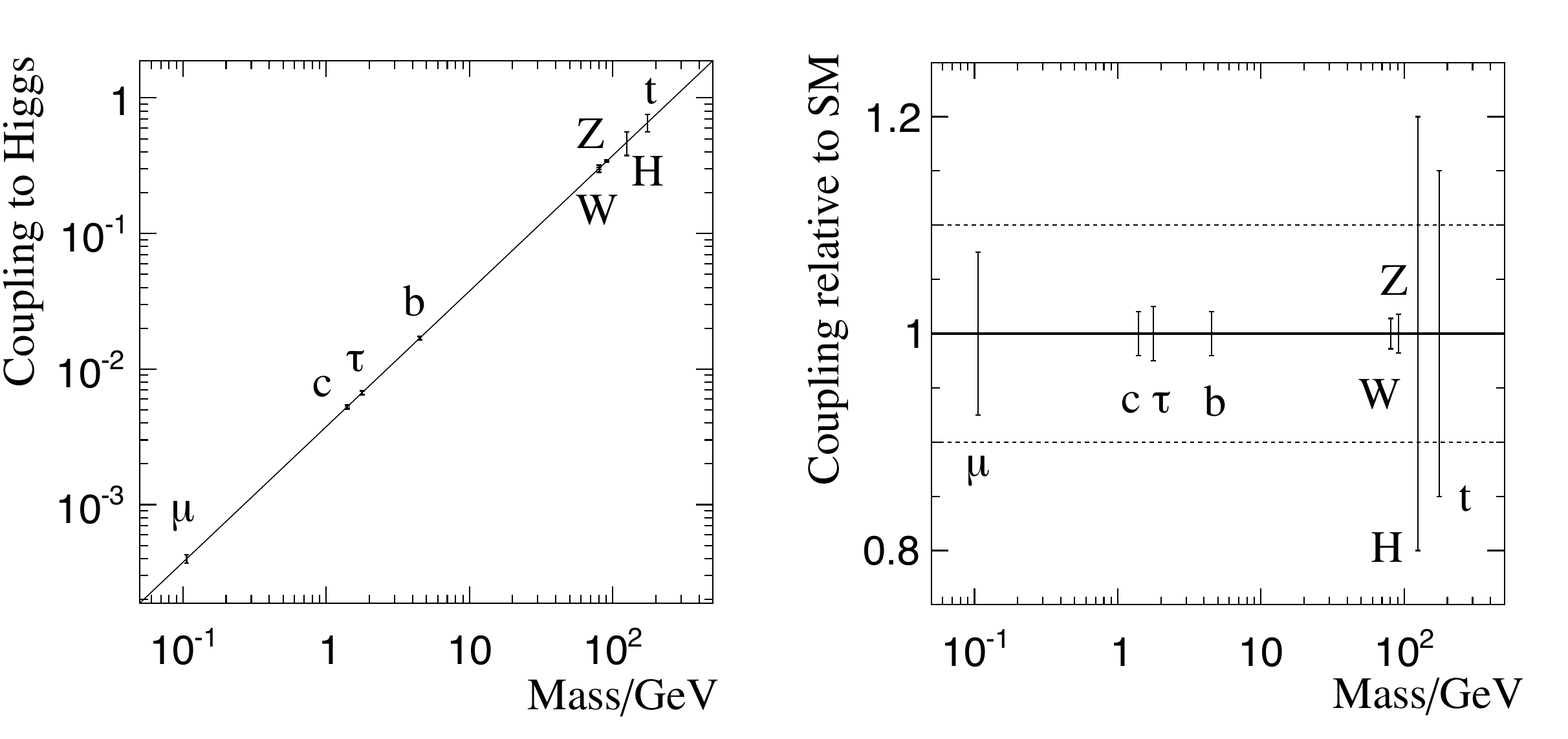}
\caption{An illustration of the typical precisions to which the relation between the Higgs couplings to
the masses of the particles can be tested at a linear collider,  assuming operation at one
energy point  below and one above $\roots=500\,\GeV$ with the integrated luminosities 
of Table~\ref{tab:higgs:events}. The ultimate sensitivity will depend on
the precise integrated luminosity recorded and the centre-of-mass energies at which the LC is operated.
The two plots show the absolute and relative precision that can be reached. The values shown assume SM couplings.
 \label{fig:higgs:couplingrel}}
\end{center}
\end{figure}

Whilst the precise measurements at a LC of the Higgs couplings to gauge bosons, fermions and to itself 
are of interest in their own right, they will be crucial 
for testing the fundamental prediction of the Higgs mechanism
that the Higgs coupling to
different particles is proportional to masses, as summarised in Figure~\ref{fig:higgs:couplingrel}.

\begin{figure}[t]
\begin{center}
\raisebox{0.5cm}{\includegraphics[width=0.5\textwidth]{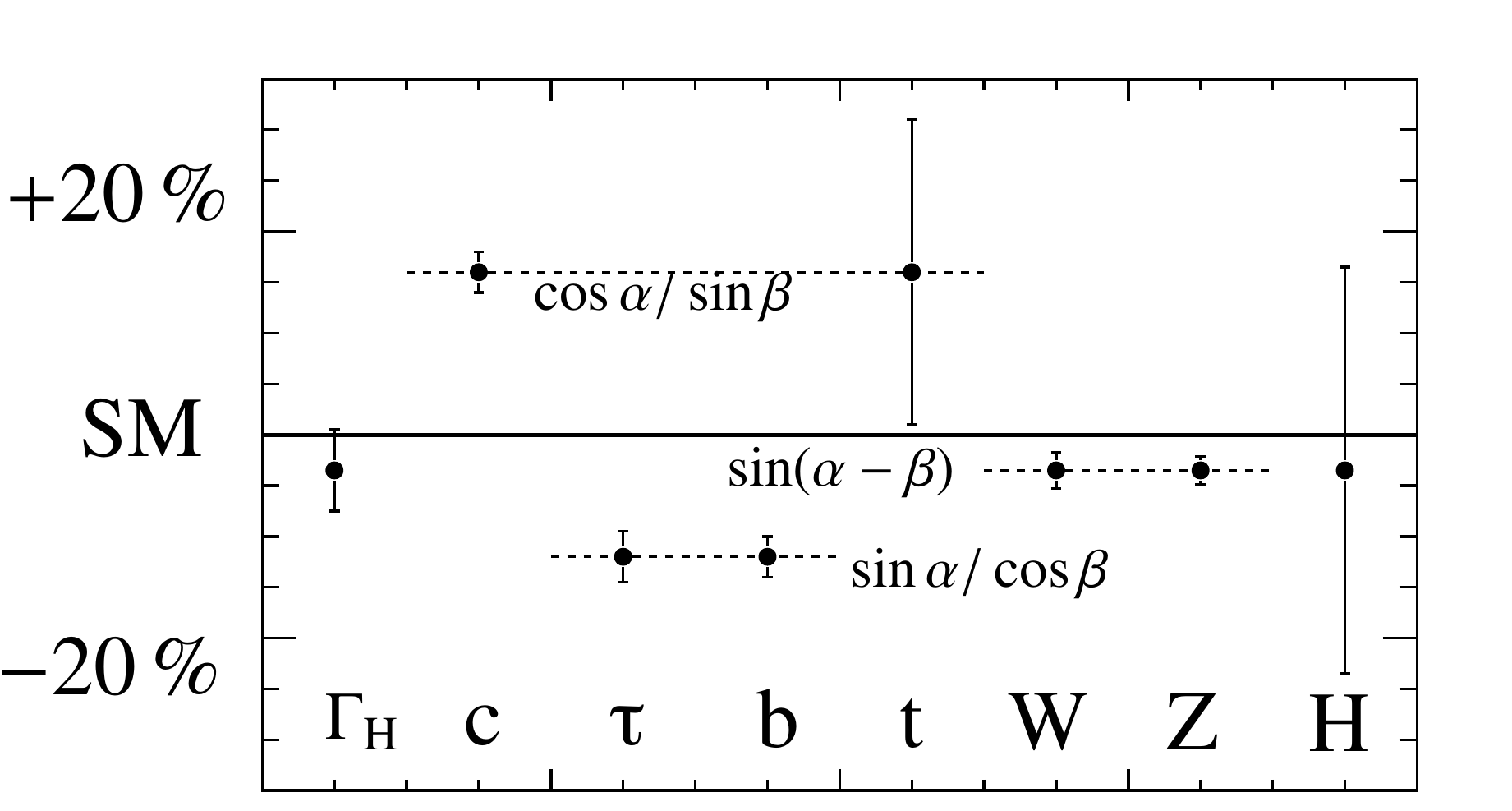}}
\includegraphics[width=0.37\textwidth]{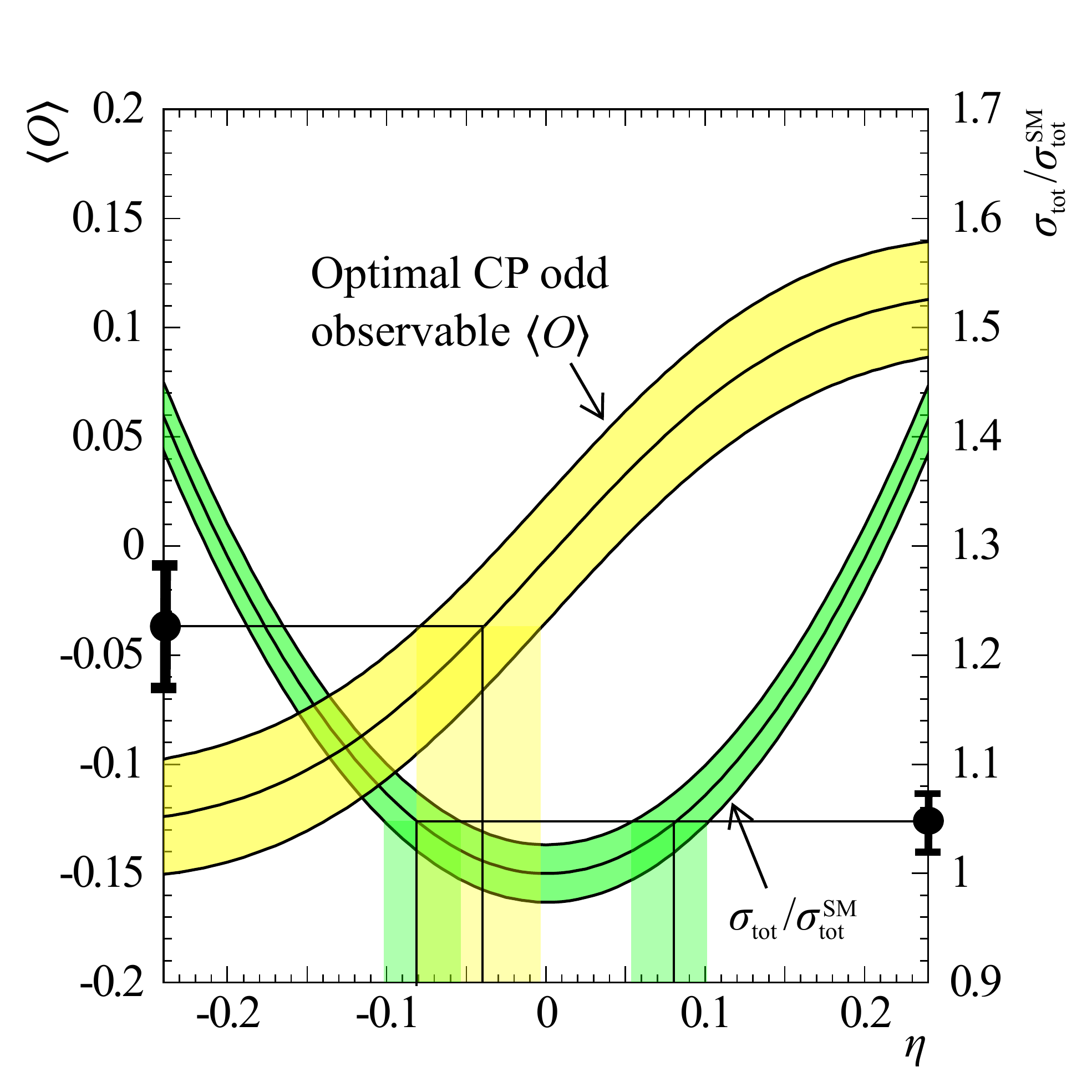}
\caption{Left: Typical deviations of the Higgs couplings to different
particles from the SM predictions in a Two-Higgs-Doublet model. The LC 
precisions for the various couplings are the same as in
Figure~\ref{fig:higgs:couplingrel}.
Right: Determination of the admixture $\eta$ of a CP-odd state in $\epem \to \Zzero\Higgs$
at $\sqrt{s} = 350$~GeV with 500~fb$^{-1}$, using the measurement of the
cross section together with an `optimally chosen' CP-odd observable.
 \label{fig:higgs:couplingbsm}}
\end{center}
\end{figure}

The precise measurements at a LC will provide
a powerful probe of the structure of the Higgs sector. The SM with a single Higgs doublet 
is only one of many possibilities. The model-independent measurements at a LC will
be crucial to distinguish between the different possible manifestations of the underlying physics. 
It is a general property of many extended Higgs theories that the
lightest Higgs scalar can have nearly identical properties 
to the SM Higgs boson. In this so-called decoupling limit, additional states of
the Higgs sector are heavy and may be difficult to detect both at the LHC and LC.
Thus, precision measurements are crucial in order to distinguish 
the simple Higgs sector of the SM from a more complicated scalar sector. 
Deviations from the SM can arise from an extended structure of the Higgs sector, for
instance if there is more than one Higgs doublet. Another source of
possible deviations from the SM Higgs properties are loop effects from BSM
particles. The potential for deciphering the
physics of EWSB is directly related to the sensitivity for verifying deviations from the SM.
For example, in Figure~\ref{fig:higgs:couplingbsm} (left) the typical
deviations from the SM predictions for a Two-Higgs-Doublet
model are compared to the precision on the couplings achievable at a LC.  
In this example, the high-precision measurements at the LC would
clearly indicate the non-SM nature of the EWSB sector.
 
Furthermore, small deviations from SM-like behaviour can
arise as a consequence of fundamentally different physics of EWSB. For example, if an additional fundamental 
scalar such as the radion mixes slightly with the Higgs boson, the subtle shifts compared to the SM Higgs boson in 
the branching ratios and overall decay width may only be discernible through
the high-precision and model-independent measurements of couplings available at a LC.

\subsection{Higgs Boson Mass, Spin and CP Properties}

A LC is the ideal place to measure the properties of the Higgs boson. For example,
the mass of the Higgs boson can be determined at a LC with a precision of
better than 50\,MeV, either from the recoil mass distribution at  $\sqrt{s} = 250$\,GeV or 
from the direct reconstruction of its decay products. This would 
improve  on the precise measurement obtained from the $\photon \photon$ decay mode at the LHC.

The spin of the Higgs boson can be obtained through 
the Higgs-strahlung process from the threshold dependence of the 
cross section as well as angular distributions of the $\Zzero$ and its decay 
products. For example, 
a threshold scan with an integrated luminosity of just $20\,\text{fb}^{-1}$ at each point is sufficient 
to establish the spin of the Higgs boson. Although the determination of the Higgs boson spin will be achieved early 
at the LHC, a LC provides a unique window into the possibility of detecting
CP violation in the Higgs and top sector.
Furthermore, the energy dependence of the Higgs-strahlung cross section in the SM 
contains a factor $\beta$, whereas for a CP-odd Higgs boson with $J^{\rm PC} = 0^{+-}$,
the corresponding factor would be $\beta^3$. Again the threshold behaviour of the cross section
can differentiate between the two spin-0 cases.  


Angular correlations in $\epem \rightarrow \Higgs\Zzero \rightarrow 4\ferm$
as well as $\Higgs \rightarrow \tauon^+\tauon^-$ decays are also sensitive to
the CP nature of the Higgs state. 
Since  {\it a priori}  the observed Higgs state can be an admixture of
CP even and CP odd states, the determination of the CP properties is experimentally more
challenging than the measurement of the spin of the Higgs boson.
For a Higgs boson, $H$, the most general model
independent expression for the $H VV$  vertex can be written as 
\begin{equation*}
g_{H VV} =  -{g}{M_V}\left[ \alpha g^{\mu\nu} +
\beta \left(p \cdot q\, g_{\mu\nu}/M_V^2 - p_\nu q_\mu  \right)
+ i\, \gamma /M_V^2\epsilon_{\mu\nu\rho\sigma}p^\rho q^\sigma
\right],
\end{equation*}
where $V$ represents either a $\Wboson$ or $\Zzero$ boson and $p,q$ are the four momenta of the two vector bosons. 
For a SM Higgs $\alpha=1$ and $\beta=\gamma=0$. In contrast,
for a pure CP odd Higgs boson, $\alpha = \beta = 0$, and $\gamma$ is
expected to be small.
A LC provides a unique laboratory to determine $\alpha$, $\beta$ and $\gamma$ and probe the complete tensor structure of the $\Higgs VV$ coupling 
and the CP properties of the Higgs boson.
For example, it has been shown that angular observables  can be used to measure 
the mixing between a CP-even and a CP-odd Higgs state to an accuracy of 
$3-4\%$~\cite{AguilarSaavedra:2001rg}; see Figure~\ref{fig:higgs:couplingbsm} (right).
The measurements of the CP properties of the Higgs based on the $\Higgs VV$ coupling, both at the LHC and a LC, project out the CP-even
component of the Higgs and therefore require very large luminosities. A LC is unique in that the measurement of the threshold behaviour of the process 
$\epem\rightarrow \Qt \AQt \Higgs$, which depends on the $\Higgs \ferm\Aferm$ coupling, provides an unambiguous determination of the CP 
of the Higgs boson and provides the potential for a precision measurement of CP-mixing, even when it is small.

\section{Top  and the Gauge Sector}

In addition to the precision studies in the Higgs sector, a further important part of the programme is
establishing the detailed profile of the top quark and studying  the gauge sector with high precision, 
to probe the dynamics of EWSB and BSM physics.

\subsection{Top Physics}

The top quark plays a very special role in the SM. Being the heaviest of the 
fundamental fermions it is the most strongly coupled to the EWSB sector and 
hence intimately related to the dynamics behind the symmetry
breaking mechanism. Its large mass affects the prediction for many SM parameters, including 
the Higgs mass and the $\Wboson$ and  $\Zzero$ couplings, through radiative corrections.
High-precision measurements of the properties and interactions of the top quark
can have sensitivity to physics at mass scales much above the EWSB
scale. These studies are therefore a very important laboratory for explorations of 
SM and BSM physics. A LC will have broad capabilities to establish
the top-quark profile in a precise and model-independent way.

\medskip\noindent {\it Top Quark Observables}

The top mass measurement at the Tevatron has reached an accuracy of about 1~GeV. 
While the statistics at the LHC will be huge, because of (theoretical) systematic effects, it 
appears nevertheless questionable whether a further significant improvement of this measurement can be reached. 
In particular, an important systematic uncertainty is
associated with the problem of how to relate the  mass parameter  that is
actually measured at the Tevatron and the LHC
to a parameter that is well-defined so that it can be used as an input for
theoretical predictions in the SM (or its extensions), such as the
$\overline{\rm MS}$ mass. The relation between those parameters is
affected by non-perturbative contributions, which can be the limiting
factor in further improving the accuracy of the top-quark mass from
measurements at hadron colliders. At the LC the measurement of
the top-quark mass from the $\Qt\AQt$ threshold will be unique
since it will enable a high-precision measurement of a ``threshold
mass'', for which the relation to a well-defined top-quark mass is precisely known and theoretically well under control.

The statistical precision 
from a threshold scan (see left panel of Figure~\ref{fig:CLICmAReach} for an example)
at the LC with approximately $30\,\text{fb}^{-1}$ 
will be about 20\,MeV for the top-quark mass and 30\,MeV for the top width. 
Including the systematic uncertainty from relating the ``threshold
mass'' to the suitable mass parameter of the SM yields an overall
precision on $m_{\Qt}$ of better than 100\,MeV, which corresponds to an
order of magnitude improvement compared to the measurement at hadron
colliders. 
\medskip\noindent
{\it Top-antitop asymmetries}

Besides the measurements of the top-quark mass and width, the top
physics programme at the LC offers a variety of further
observables that have a high sensitivity to potential effects of new
physics. Some interesting  examples are the forward-backward asymmetry in top-antitop 
production, $A_\text{FB}$ , the beam polarisation asymmetry $A_{\text{LR}}$ and 
the polarisation of the top. 
The first of these, $A_{\text{FB}}$,  has received a lot of attention lately.
Both CDF and D0 experiments have reported a possible deviation of this asymmetry 
from the SM prediction in $\proton \Aproton$ collisions whereas the measurements of a related
asymmetry for the  $\proton\proton$ initial state at the LHC currently show no significant 
deviation from the SM  prediction. 
Due to the clean LC environment, one expects a significant improvement with respect to Tevatron and LHC measurements.
Accuracies of about $5\%$ can be achieved, which can probe, for example, Kaluza-Klein excitations of the gluons up to 10-20 TeV.

\medskip\noindent{\it Couplings to Gauge Bosons}

Precise and model-independent measurements at the LC of the top couplings to
weak gauge bosons will be sensitive to BSM sources~\cite{Djouadi:2007ik}. 
The production of $\Qt\AQt$ pairs in $\epem$ collisions and the subsequent
decay of the top provide a sensitive probe of the $\Qt \AQt \photon$ and $\Qt \AQt \Zzero$ vertices. Since the top quark decays before it hadronises, 
not just the cross-sections and angular distribution of the
produced top, but also various angular distributions of the
decay products of the top, which retain the memory of its polarisation,
can be used effectively towards this end. 


A study of $\epem \rightarrow \Qt \AQt \rightarrow \ell^\pm + {\rm jets}$ can lead to 
sensitivity below the percent level for BSM correction terms to $\Qt\AQt\Zzero$ and $\Qt\AQt\photon$ vertices at $\sqrt{s}=500\gev$ and $\sim 100$ fb$^{-1}$~\cite{AguilarSaavedra:2001rg}.
Use of polarised beams and polarisation asymmetries can 
improve matters by providing observables that can disentangle different couplings and
also increase the accuracy at a given luminosity.

Measurement of the $\Qt \AQt$  production below
threshold, assuming that the top width is 
measured to the above-mentioned accuracy, will allow a measurement of $g_{tbW}$ at the few percent level.
With  such precision, 
a variety of new physics models such as Little Higgs Model or models of top flavour~\cite{Djouadi:2007ik} can be probed, for example, with simultaneous measurements of the 
$\Qt \AQt \Zzero$ axial  coupling and left-handed $\Qt\Qb\Wboson$ vertex.
Use of beam polarisation can even 
allow probing anomalous effects in the $\Qt \AQt \text{g}$ system, particularly by 
testing symmetries  with construction of  observables that have specific CP,T transformation properties
and are, e.g.,  T--odd, CP--even or T--odd and CP--odd. It should be noted that the LHC can give an 
indication of an anomalous $\Qt \AQt \text{g}$ coupling through a study of top-quark polarisation in top-pair production, 
but only the LC can probe the structure in an unambiguous way.
Thus the LC can map out the $\Qt$ couplings to all the gauge bosons in a precise manner 
to probe new  physics.

 \subsection{WW, ZZ Scattering  and  the Dynamics of Strong Electroweak Symmetry Breaking}
\label{sec:wwzz}

Despite the likely perturbative nature of EWSB indicated by the value of the Higgs mass, from both indirect electroweak precision
constraints and direct observation at the LHC, one point is worth remembering. 
Even with a light Higgs, there exist formulations of
EWSB, such as composite Higgs models, where the light Higgs boson is
part of a larger spectrum of strongly interacting particles, and  discernible effects of the strong dynamics are possible, affecting gauge 
boson couplings with each other. A study of $\Wboson\Wboson/\Zzero\Zzero$ scattering and $\Wboson\Wboson$  final state processes can reveal these effects.

The close connection between the 
$\Wboson\Wboson\photon$ and $\Wboson\Wboson\Zzero$ vertices and restoration of unitarity at high energies in
W pair production in $\epem$ collision means that this process is
highly sensitive to the triple-gauge-boson vertices and to heavy resonances with mass 
far exceeding the LC energy.  Further, the same 
connection underlines the importance of this measurement to look for footprints
of any new physics. The most general $\Wboson\Wboson V$ interactions consistent 
with Lorentz symmetry involve twelve (six each for the $\photon$ and the $\Zzero$)
independent couplings, out of which only four have nonzero values  in the SM.
Terms involving different couplings are characterised by different tensor 
structures and different momentum dependencies. 
Specific models of the strong dynamics have specific predictions for some of the anomalous 
couplings.

These different kinds of couplings can be disentangled from each other using 
production angle distributions and decay product angular distributions, the
latter being decided by the polarisation of the produced $\Wboson$.  High beam 
polarisations (both $\eminus$ and $\eplus$) can be used effectively to probe these.
An analysis using a fast simulation performed at the 
two energies $\sqrt{s} = 500$ GeV and $800$ GeV~\cite{Weiglein:2004hn,Djouadi:2007ik} shows that deviations
of all these couplings from their SM values can be measured to better than
one per mil  level with luminosities 
up to $1$ ab$^{-1}$. In many cases the measurements 
are up to an order of magnitude better than  the capabilities of a 14 TeV LHC 
that have been projected~\cite{Weiglein:2004hn,Djouadi:2007ik}. 

A chiral Lagrangian for EWSB has numerous operators that govern the interactions of the vector boson degrees of freedom. 
For example in the Minimal Strong Coupling Theory a small correction term, absent in the SM, 
yields a measurable contribution to the anomalous magnetic moment of the W boson.
While it is marginally measurable at the LHC, it is readily observable at a 500 GeV LC~\cite{Abe:2001wn}.

The above is an example of deviations in the triple-gauge-boson vertices due to strong dynamics in the EWSB sector. 
There are also deviations in quartic boson interactions, which directly affect pure gauge boson scattering through 
local contact interactions, such as $\Wboson\Wboson\to \Wboson\Wboson$. 
The processes
 $\epem\to \nue\Anue \Wboson^+\Wboson^- \to \nue\Anue jjjj$ and  
$\epem\to\nue\Anue \Zzero\Zzero \to \nue\Anue jjjj$ have been studied for LC at $\sqrt{s}=1\, {\rm TeV}$ with $1\, {\rm ab}^{-1}$ 
of integrated luminosity~\cite{Abe:2010aa}, with a view to study these anomalous quartic vertices. The LC sensitivity is comparable to
the values predicted in models of strong dynamics in the EW sector, where the non-SM operators are constrained  to be consistent with 
the EW precision tests.
These measurements require study of angular correlations among the decay products of the $\Wboson/\Zzero$ and further needs  separation of the
$\Wboson$ and $\Zzero$ final states decaying to quarks.  This indeed has been a benchmark requirement, which has driven the need for excellent 
jet-energy resolution, which in turn has driven the design of LC detector concepts and has been shown to be achievable.

As mentioned above, one could have strong dynamics at the origin of EWSB, even for a light Higgs boson, and it could be a composite particle remnant. In the case of these composite 
Higgs models, the Lagrangian of the Higgs boson interactions with the vector bosons receives correction terms proportional to $v^2/\Lambda^2_{\rm comp}$, where $\Lambda_{\rm comp}$ is the compositeness scale.
Precision measurements of production cross-sections $VV\to VV$, $VV\to \Higgs\Higgs$, and $\epem\to \Higgs\Zzero$ 
provide sensitivity to the composite scale. The results show that 14 TeV LHC with $100\, {\rm fb}^{-1}$ of integrated luminosity 
should have sensitivity of $\Lambda_{\rm comp}$ up to $7\, {\rm TeV}$, $500\, {\rm GeV}$ LC with $1\, {\rm ab}^{-1}$ up to 
$45\, {\rm TeV}$, and $3\, {\rm TeV}$ LC with $1\, {\rm ab}^{-1}$ up to $60\, {\rm TeV}$~\cite{Linssen:2012hp}.

\section{New Physics}
\label{section:newphys}

The physics programme of the LC for exploring Terascale physics consists
of three broad categories, all of which will be crucial for revealing
the possible structure of new physics and for
discriminating between different possible manifestations of
physics beyond the SM:

\begin{itemize}

\item
{\it Refining LHC discoveries: }
Phenomena of new physics  discovered at the LHC 
will be probed at the LC in a clean experimental environment and with high
precision. This is expected to be decisive for revealing the physics mechanisms
behind the observed phenomena.

\item
{\it New direct discoveries:}
The LC will have a potential for direct discoveries that is
complementary to the LHC. In particular, the searches for
colour-neutral states of new physics, including the full structure of the Higgs sector, 
will have a discovery potential that far surpasses that of the LHC.

\item
{\it Discoveries through precision:}
Measurements of observables involving known particles at
the LC with the highest possible precision will have a high sensitivity
to resolving the fingerprints of new physics, which in many scenarios
only manifest themselves in tiny deviations from the SM prediction. 
Examples for the achievable precisions can be found on page 230 of~\cite{Linssen:2012hp}
and Table 5.2 of~\cite{Djouadi:2007ik}.

\end{itemize}

In the following subsections we give examples of new physics where one
or more of the above categories of the LC physics programme is on
display, some examples of the last having been presented in the earlier discussions
of the precision studies in the Higgs, top and the gauge sector.

\subsection{New Electroweak Matter States}

In the BSM context, there are many electroweak states that are
well known to be difficult to find directly at the LHC.   The event rates at the LHC are small in comparison to strongly interacting particle creation that makes for a challenging background environment. 

Of the many ideas that one can use to demonstrate how well new electroweak matter states can be found at a LC, perhaps the most well known is supersymmetry. Supersymmetry provides a good study ground not only because it is a highly motivated scenario for physics beyond the SM, but also because it provides a rather complete and calculable framework beyond the SM  with multiple new scalars and fermions of different gauge charges.  

The LHC has very good prospects for discovering
pair-produced coloured particles up to masses of 2--3~TeV. 
On the other hand, 
non-coloured particles, charginos, neutralinos and sleptons are not copiously produced by the LHC. 
Although these electroweak particles may be found in cascade decays 
of strongly interacting squarks and gluinos, their prospects for discovery rely on the details of the model. Their accessibility through the decay chains is unlikely to be complete. On the other hand, an $\epem$ collider running at sufficiently high centre-of-mass energy potentially can produce
each of these states directly with manageable backgrounds leading to
discovery.  The discovery reach for these particles produced in pairs at
the LC is usually close to $\sqrt{s}/2$, and in some cases even higher if $m_A\neq m_B$ in $\epem\to AB$ searches. 

\begin{figure}[t]
\begin{center}
\includegraphics[width=0.39\linewidth]{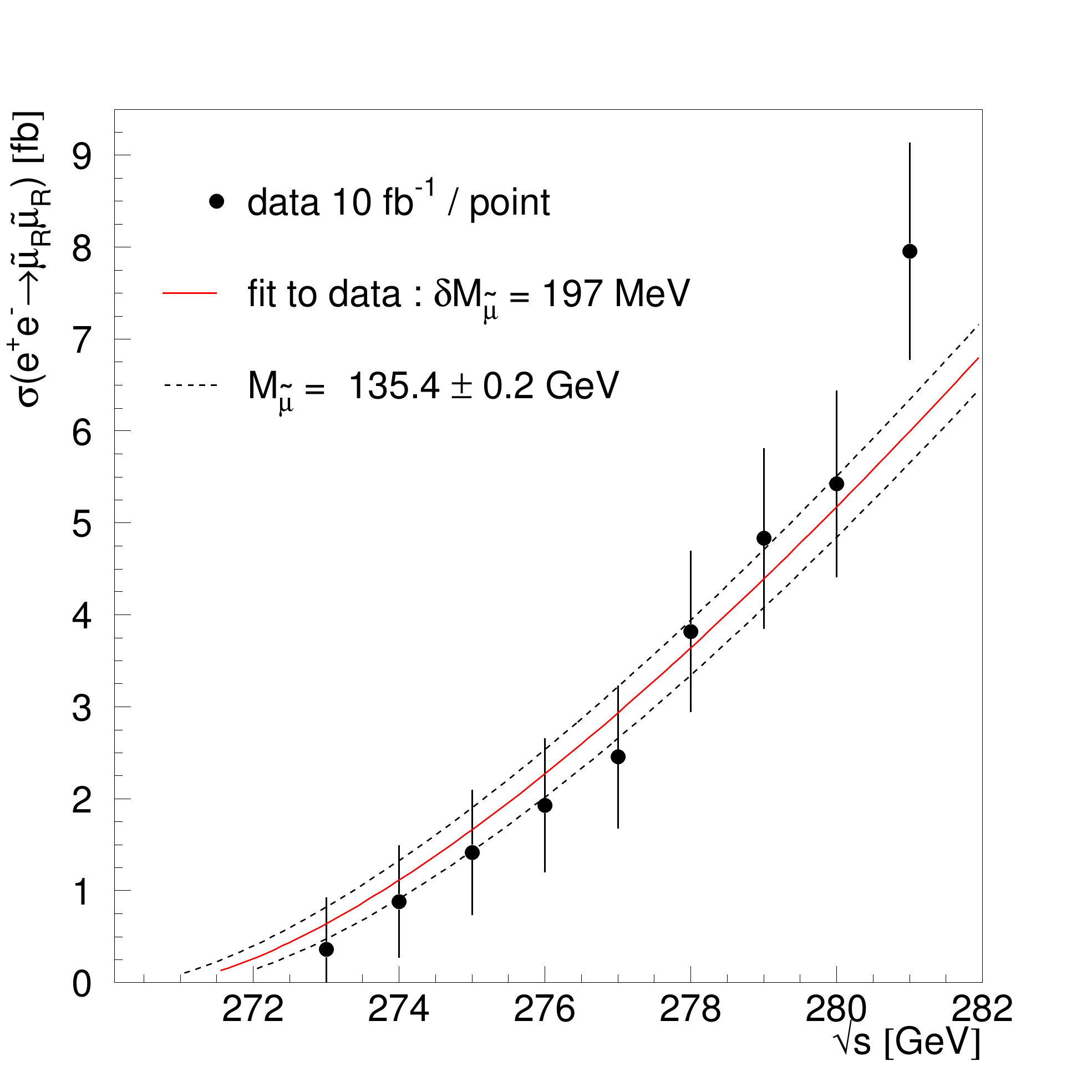}
 \includegraphics[width=0.40\linewidth]{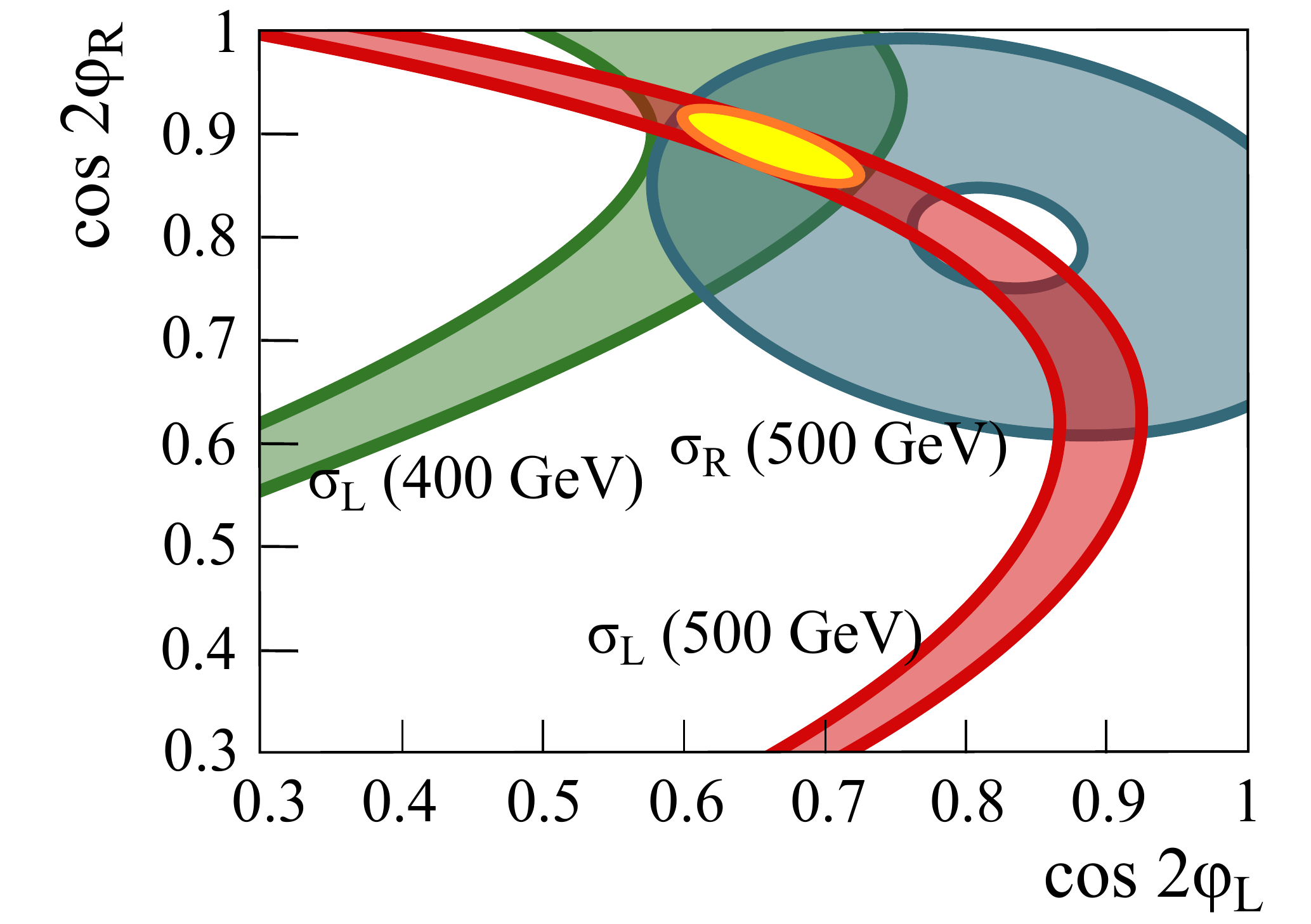}
 \caption{Left: Cross section at threshold for the production of the
superpartners of the right-handed muons at the LC, 
$\epem \to \tilde \muon_{\rm R}\tilde \muon_{\rm R}$, from which the spin
of the produced particles can be determined and their mass can be
precisely measured (limited by statistics; the plot shows a `difficult' scenario with backgrounds from other light SUSY particles). Right: Determination of the chargino mixing angles
$\cos 2\phi_{\rm L,R}$ from LC measurements with polarised beams and at
different centre-of-mass energies. 
 \label{fig:threshold-mixing}}
\end{center}
\end{figure}

The precision studies that are possible at a LC can test many of
the properties of the discovered particles, such as per mil precise
values of their masses and their couplings to SM particles, and
assignment of spins. This can be accomplished through several means,
including collecting high integrated luminosity at high energies and also
through threshold scans, which are particular good at measuring the
spin due to the shape of the cross-section versus near-threshold energy.
The precise measurement of
the couplings then enables tests and resolutions of the underlying
structure.
Detailed measurements of this kind will be crucial for discriminating different sources of new physics.
For example, the predictions for the spins, quantum numbers, couplings and certain mass relations are characteristic features of supersymmetry
that need to be experimentally tested.
Furthermore, the precision
measurements of the electroweak superpartner masses at the LC, combined
with the measurements of the masses of the strongly interacting
superpartner masses at the LHC, enable us to test many ideas of the
underlying organisational principle for supersymmetry breaking. 
Through renormalisation group scaling of well-measured parameters one
gets access to the high-scale (e.g., scale of Grand Unification $\sim 10^{16}\gev$) structure of the theory, enabling a test
of properties like coupling and mass unification.

\subsection{Dark Matter}

It is well established now that the Universe must contain a sizable
fraction of cold dark matter.  An ideal candidate for this dark matter
is a chargeless massive state $\chi$ that interacts with approximately
weak gauge force strength (weakly interacting massive particle,
``WIMP'').  

There are several model-dependent prospects for finding dark matter at the LHC and LC. These include cascade decays of parent particles  
that terminate in a stable dark matter particle candidate that carries off missing energy. These missing energy signature rates depend crucially on many different parameters of the overarching theory and generally have little to do with the couplings directly relevant to the dark matter particle itself. 

On the other hand, a more direct and less model-dependent search for
dark matter focusses on the (effective) $\ferm\Aferm\chi\chi$ interaction.
If the annihilation cross-section is in accordance with the observed relic density, 
there are good prospects for the production of dark matter directly at colliders through 
$\ferm\Aferm \to \chi\chi\photon$, where the initial-state radiated photon (or gluon) is needed to tag the event.
The sensitivity of this process at the LHC is limited because of
significant backgrounds.   
While at the LHC and in direct detection searches the WIMP interaction with quarks is 
probed, the LC provides complementary information on the WIMP interaction with electrons.
Within the clean LC environment, making use of polarised beams, the WIMP mass, the strength and the chiral structure 
of the $\epem\chi\chi$ interaction, as well as the dominant partial wave of the
production process can be determined. 

LC measurements can also provide a comprehensive set of high-precision 
experimental information on the properties of the dark matter particle and 
the other
states affecting annihilation and co-annihilation of the dark matter
particle. This can then be used to predict the dark matter relic
density in our Universe. The comparison of the prediction based on the
measurements of new physics states at the LHC and the LC with the 
precise measurement of the relic density from cosmological data would
constitute an excellent test of the dark matter hypothesis.

\subsection{Additional Higgs Bosons}

After the confirmation of the existence of a state compatible with the SM Higgs boson, there is still the prospect
of additional Higgs bosons in the spectrum.  These additional Higgs bosons include extra singlet Higgs bosons that mix with the SM-type Higgs boson. Or, there may be an extra ${\rm SU(2)}_{\rm L}$ doublet that fills out the full Higgs sector of the theory.

\begin{figure}[t]
\begin{center}
\includegraphics[width=0.45\linewidth]{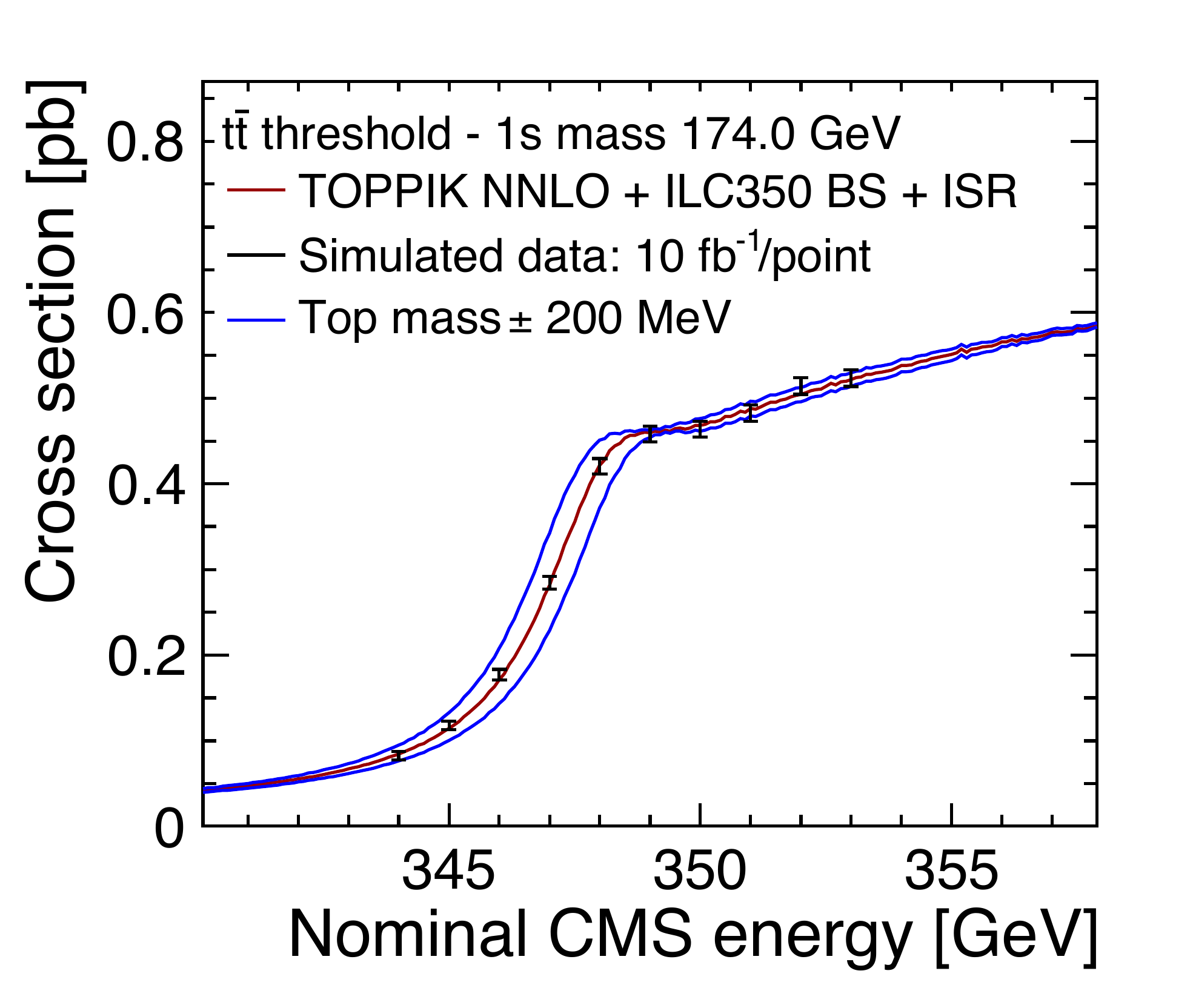}
\raisebox{0.7cm}{\includegraphics[width=0.45\linewidth]{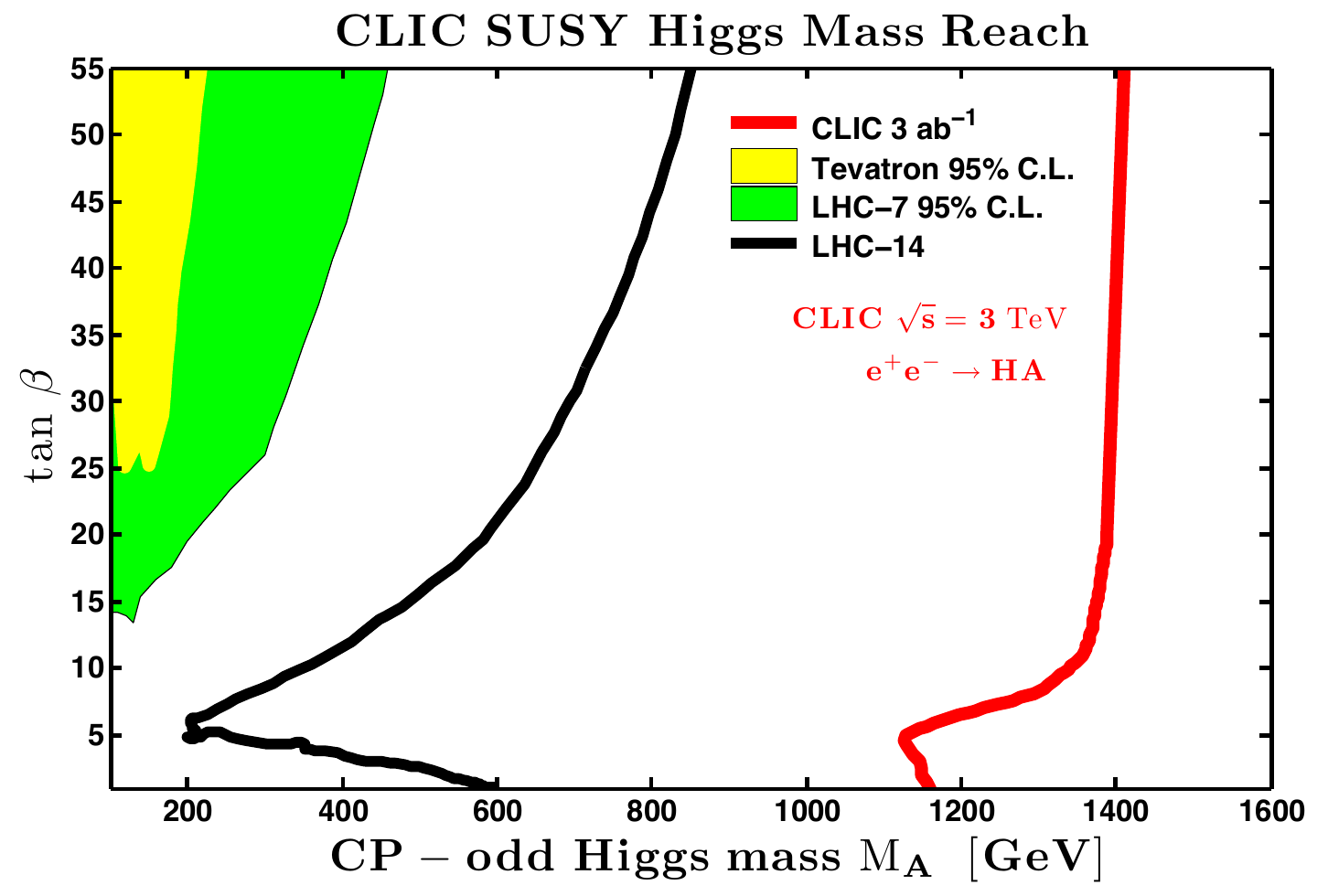}}
 \caption{Left: 
 The $t\bar{t}$ production cross-section scan near the threshold, leading to $30$~MeV determination of the top mass.
 The study is based on full simulation of the ILD detector and includes initial state radiation, beamstrahlung and other machine-induced effects~\cite{Barklow:2012}.
Right: Search reach in the $m_A-\tan\beta$ plane for LHC and for $3\tev$ LC.
The yellow and green regions are limits already in place from Tevatron
and LHC (7 TeV run) analyses. The black line is a $5\sigma$ discovery projection for the LHC
at 14 TeV with $300\, {\rm fb}^{-1}$~\cite{Gianotti:2005} (limits are roughly $150\, {\rm GeV}$ uniformly higher with $3000\, {\rm fb}^{-1}$), 
and the red line is a projection for 
$3\, {\rm TeV}$ $\epem$ with $3\, {\rm ab}^{-1}$ of integrated luminosity~\cite{Linssen:2012hp}.   
 \label{fig:CLICmAReach}}
\end{center}
\end{figure}

Again, supersymmetry provides a calculable framework through
which to analyze the discovery prospects of an extra Higgs boson. Over
a large part of the parameter space the Higgs sector consists of one
light state ($m_\higgs\lsim 135\, {\rm GeV}$) whose couplings are very
similar to the SM Higgs boson, and four extra states ($\text{A}^0$, $\Higgs^0$ and
$\Higgs^\pm$)  of nearly equal mass.  Figure~\ref{fig:CLICmAReach} shows the direct discovery reach of the heavy Higgs bosons at the LHC and a 3 TeV LC as a function of $m_A$.   The result is impressive, with a search capacity for the heavy Higgs near $\sqrt{s}/2$ for the LC.
If the dark matter particle has less than half the mass of a Higgs boson, invisible Higgs decays could be another good way to identify it. 
This possibility can be studied in detail at the LC for all Higgs bosons within its kinematic reach.

An extended Higgs sector could also contain a light Higgs, possibly in
addition to a SM-like Higgs at about 125 GeV, with a mass below the LEP
limit of about 114 GeV and with suppressed couplings to gauge bosons.
While at the LHC the search for such a light Higgs
state will be very challenging in the standard search channels, at the LC there will be a high
sensitivity for probing scenarios of this kind. 

\subsection{New Gauge Boson Interactions}

The quintessential example of a new gauge boson is a $\Zzero^\prime$ boson. The mass reach for direct discovery at the LHC of an ``ordinary" $\Zzero^\prime$ boson, whose couplings to the SM fermions are ${\cal O}(1)$, is generally about $5\tev$.
However, it is well documented that through non-resonance observables an
$\epem$ collider with energy above a few hundred GeV has an even higher
reach for detecting BSM signals. This is accomplished by studying precisely the observables of the $\epem\to
\ferm\Aferm$ processes. Small deviations in $\sigma^{\ferm\Aferm}_\text{tot}$, $A_{\text{FB}}^\ferm$ and
$A^\ferm_\text{LR}$ can be found for $\Zzero^\prime$ masses well above the centre-of-mass
energy of the machine. For example, at a 500\,GeV LC with $1\,
{\rm ab}^{-1}$ of integrated luminosity, a BSM signal is
detectable in the left-right model (i.e., theory with ${\rm SU(2)}_{\rm R}\times {\rm U(1)}_{\rm B-L}\to {\rm U(1)}_{\rm Y}$) if the corresponding $\Zzero^\prime$ has a mass below 9\,TeV, which is more than one order of magnitude beyond
the centre-of-mass energy of the collider. This search reach increases
to about 16\,TeV at a 1\,TeV LC (see sec.~5.2.1
of~\cite{AguilarSaavedra:2001rg}) and to
well beyond $30\, {\rm TeV}$ for a 3\,TeV LC (see sec.~1.5 of~\cite{Linssen:2012hp}).

\subsection{Model-Independent Searches}

Some of the discussion above has revolved around specific model
scenarios. However, it must be emphasised that the LC is an excellent
machine to do model-independent analyses in the context of the uniquely clean
$\epem$ collision environment. Searches can be made to test whether 
the event rates in different channels are anomalous, and thus indicate
the presence of new physics. 
A minimum number of theoretical assumptions are necessary to determine the
spin, mass and couplings of new particles, which can then be used in a second step
to obtain theoretical interpretations in different models. Thus, instead of referring to a
particular class of models, like the discussion above of $\Zzero^\prime$ effects
suppressed by $M_{\Zzero^\prime}$, one can also interpret the LC results in terms
of general effective operators, such as
non-renormalizable contact operators
suppressed by a scale $\Lambda$.  These more general interpretations of the
LC sensitivities may not always be stated explicitly since many studies have been 
carried out within a well-defined BSM model, but it is an
advantageous feature of the LC that such model-independent
interpretations are possible.

With the so-called GigaZ option of the LC, i.e.\ a run at the $\Zzero$ peak with polarised 
$\eminus$ and $\eplus$ beams collecting about $10^9$ events, the LC can
provide high-precision measurements that
have a very high sensitivity to effects of new physics, which are probed
in a model-independent way.
In particular, the GigaZ run would reduce the present experimental 
uncertainties on the effective weak mixing angle, $\sin^2\theta_W^{\rm eff}$, 
by more than an order of magnitude, 
and resolve or confirm the significant ($3\sigma$) disagreement 
between the two most precise determinations of $\sin^2\theta_W^{\rm eff}$ 
from $A_{\rm FB}^\Qb$ at LEP and $A_{\rm LR}$ at SLC.
As an example, the precision achievable for $\sin^2\theta_W^{\rm eff}$ at
GigaZ has the potential to 
reveal the impact of new physics even in a scenario where no states of
physics beyond the SM would be observed at the LHC and
the first phase of a LC.

\section*{Executive Summary}

The observation at the LHC of a SM-like Higgs particle provides the first direct test of the minimal
SM EWSB scenario 
of a single scalar doublet Higgs field producing the vacuum expectation value. This discovery makes 
the physics case for a LC extraordinarily strong. 
The LC provides the capability to study the details of this new form of matter, 
establishing agreement with the SM predictions to new levels of sensitivity,
or revealing a break from the patterns expected in the SM.
The precision of the LC opens sensitivity to new physics well beyond
the LC's direct reach, enabling detection before discovery, such as past indirect evidence for the
Higgs boson, the top quark, the charm quark, and the weak gauge bosons.  

The most powerful and unique property of the LC is its flexibility.
It can be tuned to well-defined initial states, including polarisation, allowing numerous model-independent measurements, from the Higgs threshold to multi-TeV operation, as well as the possibility of 
unprecedented precision at the Z-pole (GigaZ).
Furthermore, the relative simplicity of the production processes and final-state configurations
makes complete and extremely accurate reconstruction and measurement possible.
The envisioned physics programme includes precision measurements of many Higgs decay widths, some of which are uniquely accessible at the LC ($\Qc\AQc$, $\text{g}\text{g}$, 
the invisible mode and the full width), 
decisive tests of the CP properties of the Higgs candidate,
and determinations of the top-Higgs and trilinear Higgs self couplings,
also uniquely accessible at the LC.  
Using a LC, the complete SM, including Higgs, top quark and VV interactions, can be studied, both at tree level and through quantum corrections.
The LC reaches well into new physics territory.  
Well-motivated BSM physics ideas such as dark matter, supersymmetry, composite Higgs bosons, contact interactions, and extra space dimensions could be discovered and explored.
The physics reach of the LC is essentially limited by statistics, not systematics.
Its discovery reach exceeds that
of the LHC at any integrated luminosity in many cases, and discoveries of new particles or interactions at either machine can be subjected to further precision analysis at the LC to reveal deeper structures of nature.


\vfill\eject
\bigskip\bigskip\bigskip
\section*{Addendum: Charge for the Linear Collider Report Committee}

During the international Linear Collider Workshop in Granada October 2011 it was proposed and agreed to charge a small expert group with drafting  a common Linear Collider Physics report to be submitted as input to the European Strategy process. The initiative was presented in Granada by the GDE European Regional Director (Brian Foster), the CERN Linear Collider Studies Leader (Steinar Stapnes) and the Chair of the ECFA Study for the Linear Collider (Juan Fuster), and was a result of discussions and consensus in several ILC and CLIC steering committee meetings earlier in 2011. These three subsequently suggested a composition of the expert committee based on input from the community, and proposed the mandate of the committee. The draft report has been through internal reviews, and has been made openly available to the full international LC community for further comments and suggestions before submission by end of July 2012.

\medskip
\noindent
{\it Mandate of the committee:}

The committee is requested to review the physics case for a linear electron-positron collider in the centre-of-mass energy range from around $250\, {\rm  GeV}-3\, {\rm TeV}$ in the light of LHC results up to mid-2012 and building on previous studies.
The committee should consider the case for a linear collider in terms of the physics reach
beyond that of the LHC under the assumptions in the current CERN planning; a) $300\, {\rm fb}^{-1}$ and 
b) $3000\, {\rm fb}^{-1}$.

It should assume linear collider performance based on the details contained in current documents from ILC and CLIC but without a detailed comparison of the relative performance of the machines.
The aim is to make the strongest possible case for a generic linear collider for submission to the European Strategy process.

The committee is requested to submit its draft report to the GDE European Regional Director, the CERN Linear Collider Studies Leader and the Chair of the ECFA Study for the Linear Collider by June 18th 2012.
The final version of the report should be delivered by end of July 2012.

\end{document}